\documentclass{aa}
\usepackage{graphicx,epsfig}
\usepackage{natbib}
\usepackage[usenames]{color}
\usepackage{txfonts}
\usepackage{url}
\usepackage{hyperref}

\newcommand{\kms}{km~s$^{-1}$}

\begin{document}

\title{Quiet Sun mini-CMEs activated by supergranular flows}

\author{D.E. Innes\inst{1} A. Genetelli\inst{1,2} R. Attie\inst{1} and
H.E. Potts\inst{3}}

\institute{Max-Planck Institut f\"{u}r Sonnensystemforschung,
  37191 Katlenburg-Lindau, Germany
 \and
 Universit\'e Paul Sabatier,  31062 Toulouse, France
\and Department of Physics and Astronomy, University of Glasgow,
   Glasgow, G12 8QQ, UK}

\offprints{D.E. Innes
              \email{innes@mps.mpg.de}
             }

\abstract
{The atmosphere of the quiet Sun is controlled by photospheric flows
sweeping up concentrations of  mixed polarity magnetic field.
Along supergranule boundaries and junctions, there is a strong
correlation between magnetic
flux and bright chromospheric and transition region emission.}
{The aim is to investigate the relationship between photospheric flows and
small flare-like brightenings seen in Extreme Ultraviolet images.}
{We describe observations of small eruptions seen in quiet
Sun images taken with the Extreme UltraViolet Imager (EUVI) on STEREO.
The photospheric flows during the eruption build-up phase are investigated by tracking
granules in high resolution MDI continuum images.}
{Eruptions with characteristics of small
coronal mass ejections (CMEs) occur at the junctions of supergranular cells.
The eruptions produce  brightening at the onset site, dark cloud or small
filament
ejections, and faint waves
moving with plane-of-sky speeds up to 100~\kms.
In the two examples studied, they appear to be activated by converging and rotating supergranular flows,
twisting  small concentrations of opposite polarity magnetic field.
An  estimate of the occurrence rate is about 1400 events per day
over the whole Sun. One third of these events seem to be associated with waves.
Typically, the waves last for about 30 min and
travel a distance of 80~Mm, so at any one time they cover
1/50th of the lower corona.}
{}


\titlerunning{Quiet Sun mini-CMEs}
\authorrunning{Innes et al. }

\maketitle


\section{Introduction}
Coronal mass ejections (CMEs) are large-scale eruptions of plasma into the solar
corona.
In coronagraph images, a classical CME has a three-part structure.
There is a fast outer front with a large dark cavity behind and inside a
bright dense core. Against the disk, extreme ultraviolet (EUV) images
usually detect a flare-like brightening at the onset site, a compact dark
eruption, and sometimes a
dimming propagating outwards across the disk.
CMEs are thought to be triggered by instability leading to magnetic
reconnection, and brought on by convective flows at the
footpoints of large magnetic loops. Mini-CMEs have the same characteristics
on a smaller scale. They will be most easily identified
against the quiet Sun because in active regions
they disappear into the general background activity and off-limb
there is too much line-of-sight confusion.

Mini-CMEs have not, as such, been reported previously; however it
 is likely that parts of them have. The small erupting
filaments seen by \citet{Hermans86} and \citet{Wetal00}  with a
typical lifetime of 50~min, speed of 13~\kms, and length of 20\arcsec,
may be the erupting core. If so, mini-CMEs could be expected to have the
same high rate of about 6000 events
per 24 hours on the Sun \citep{Wetal00}. They may also be related to
 the flare-like brightenings seen in high cadence TRACE 171~\AA\
quiet Sun images discussed by \citet{Ireland99}.
In that sequence, the TRACE exposure time was probably too
short to detect mini-filament eruptions as darkenings in the EUV, and the
analysis only describes brightenings.

High signal-to-noise observations
are now available at cadences fast enough to see
both mini-filament eruptions and brightenings  from
the recently launched EUV Imagers \citep[EUVI;][]{Howard08}
 on the two STEREO spacecraft. The two spacecraft have
slightly different orbits. One is inside the Earth (Ahead) and
the other outside (Behind) so that the spacecraft are gradually moving apart in order to observe
 the Sun simultaneously from different angles.
 At the time of the observations analysed here, the
STEREO spacecraft were
 12 degrees apart, and each about 6 degrees from the Earth. In principle, the
 two images can be combined to reconstruct the 3-D structure of events
 \citep{Feng07, Aetal08}. When they are near 90 degrees
  apart it will be possible
  to analyse both the on-disk and off-limb view of (mini-)CMEs.

One advantage of the small angle is that it allowed us to simultaneously observe the
photospheric magnetic field and flows in high resolution with MDI/SoHO,
 and test
the importance of supergranulation flows during the pre-eruption phase.
Convective flows below the solar surface  sweep up
small concentrations of mixed-polarity magnetic field into the distinctive
chromospheric network pattern of supergranular cells
\citep{Martin88, Wang88, Schrijver97, Hag01, Parnell02}.
The chromospheric emission and activity is
strongest along the boundaries where the magnetic flux interactions occur
\citep{Porter87, MFPS99, Detal91, Aiouaz08}.
Models of the energy released  during interaction have been suggested by several
authors \citep{Priest94, MFPS99, Longcope99, PHT02}. In these models magnetic
energy is converted into plasma heat and kinetic energy at reconnection sites,
producing hot loops and plasma jets.
Another aspect discussed by \citet{Priest94}, \citet{Moore88} and \citet{MFPS99},
and central to the observations
discussed here, is the formation and eruption of filaments above the
interacting field regions. Filaments tend to form along magnetic neutral
lines with a large degree of magnetic twist or shear
 \citep{Priest89, Balle89, Balle00, Amari00}. The twist can be measured from
 vector magnetograms, and it is caused by rotation and movement
 of the photospheric footpoints.

Most UV and EUV quiet Sun dynamic events appear to be small scale
 microflares or X-ray bright points that heat the
plasma locally. To contribute to coronal heating the energy must be
carried into the corona.
Waves generated by impulsive events have so far only been
unambiguously identified in
large CMEs and flares
\citep{Moreton60, Tetal98, KA02, Biesecker02, Warmuth04}, and these are not frequent
enough to heat the whole corona. If, however, waves are associated with mini-CMEs
and they have a rate comparable to that of mini-filament eruptions, then they
may be a viable mechanism for heating the corona.

In this paper, we show features of
small CME-like eruption events
occurring at the junctions of supergranular cells.
The EUV images  show  the
flaring hot plasma through a brightness increase, plasma evacuations due to waves
by a dimming,
 and cold plasma ejections as a darkening. We demonstrate with examples of
 space-time intensity variations, the frequency and general characteristics
 of the events. When tracking the photospheric flows and magnetic field in
  two events, we find  vortex-like structure at the supergranular
  junction preceding the eruptions.

\section{Data Preparation}
\begin{figure}
   \centering
   \includegraphics[width=9cm]{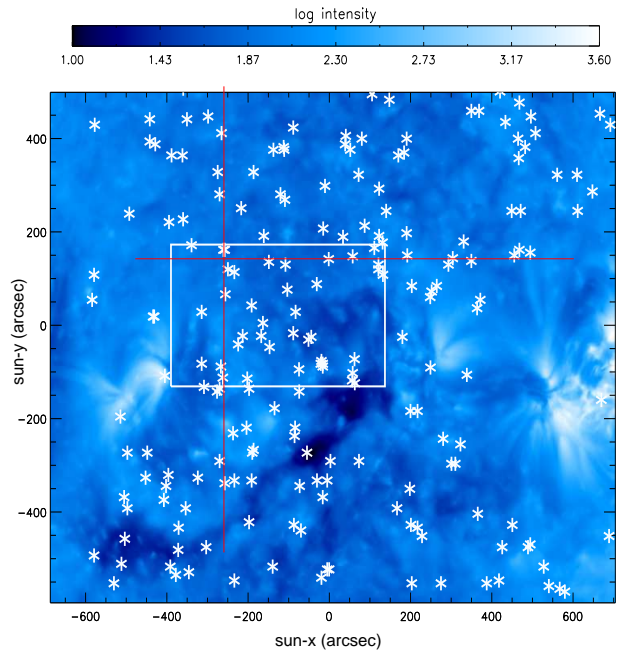}
   \caption{ The average 171~\AA\ intensity image for the
   region studied on 11 June 2007, with events and the region of high
   resolution MDI marked. The horizontal and vertical red lines indicate the
   positions of the space-time images shown in Fig.~\ref{space_time_fig}.
}
    \label{stevents_fig}
  \end{figure}

The region chosen for study (Fig.~\ref{stevents_fig})
included a small equatorial coronal hole
at disk center with  active regions on either side, and a large area
of quiet Sun to the north. It was selected because it contained a variety of
different solar structures passing disk center at a time when both
synoptic EUVI/STEREO 171~\AA\ and an 8 hour series of MDI high resolution
data were available.

EUVI/STEREO 171 \AA\ synoptic images of the full Sun were taken on
2007 June 11 with the two STEREO
spacecraft (Ahead and Behind) throughout the day
with a cadence of 2.5~min and pixel size  of 1.59\arcsec. Here only images
from the Ahead spacecraft are discussed. We have used the Behind images to
confirm faint structures seen in the Ahead images.
There
were 586 images from each spacecraft, and 195 of these overlapped with the time
of high resolution MDI observations.
The EUVI images were calibrated and rotated to the SoHO viewpoint at the time of the first
high resolution MDI image (07:51~UT) using
standard routines available in {\it SolarSoft}.

MDI high resolution (pixel size 0.625\arcsec) observations
were made from 07:51~UT to 16:06~UT with a cadence of one frame per min,
over an area 622\arcsec$\times304$\arcsec.
The continuum and magnetic field
images were de-rotated to the eastern 516\arcsec\ of the 07:51~UT image using
{\it SolarSoft} routines. The continuum
images were used to compute photospheric horizontal flows using the Balltrack
method described by \citet{Potts04, Potts07}.

The coalignment between EUVI and MDI was checked by comparing MDI full disk  with
the full disk EUVI images. Since there were the two active regions and the
limb to coalign with, we believe the coalignment to be accurate to 4\arcsec\
(two full disk MDI pixels).

\section{Eruption events}

\begin{figure}
   \includegraphics[width=8.5cm]{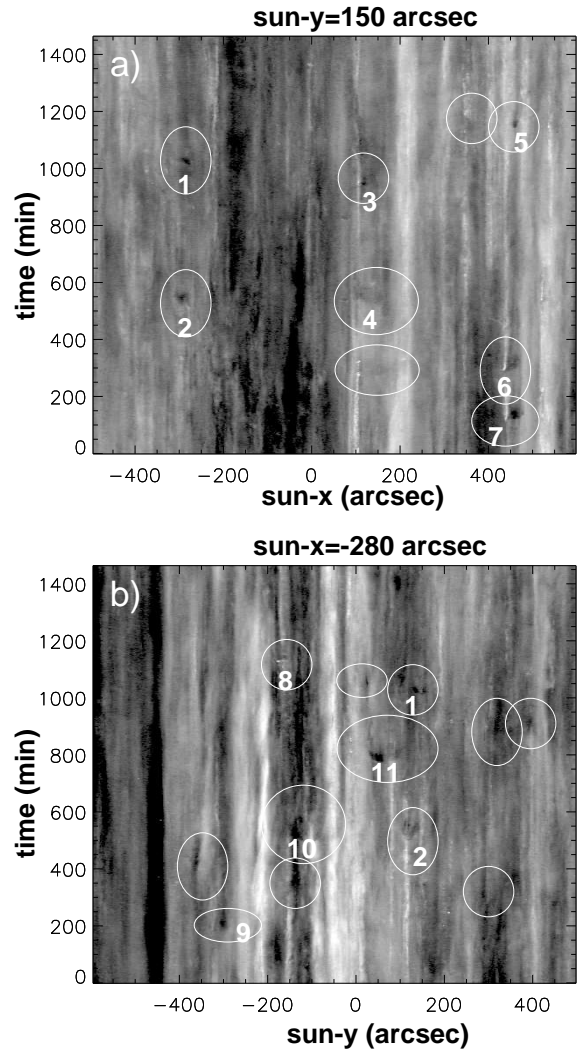}
   \caption{Time series of EUVI 171~\AA\ images for slices of quiet Sun along the
   sun-$x$ and sun-$y$ directions (see Fig.~\ref{stevents_fig}).
    Events are circled.  The slices
   cross near events 1 and 2 so these events are seen in both images. The
   numbered events are shown in more detail in Fig.~\ref{mini_events_fig}.
}
\label{space_time_fig}
  \end{figure}

Eruption events are seen with a fairly wide range of characteristics in the
EUV images.
The basic feature is a brightening at the eruption site and nearly always
a small, dark plasmoid ejection.  In stronger events there is, in addition,
 a faint wave-like feature that travels several tens
of arcsec from the source.
For this initial study, we have used a
very simple event identification method based on recognizing  emission
or absorption trains by eye in series of time-distance 171~\AA\ images.
Examples of time-distance images are shown in Fig.~\ref{space_time_fig}.
These pictures show many abrupt changes in the
EUV emission and we have had to make sensible choices for selecting events.
On the computer screen events show up more clearly, especially the faint
wave-like extensions.
The first thing we did was run through the dataset displaying space-time
images row-by-row. If an event was seen in any image, its space-time
co-ordinates were filed.
Afterwards only events seen over 6\arcsec\ (3 consecutive pixels) were kept.
Then any events closer than 40\arcsec\ in space and 50 min in time to any other recorded
event were removed. This made sure that the same event was not counted twice.
The event positions, marked as asterisks in Fig.~\ref{stevents_fig},
are those that remain.
Because there is no check on the event strength seen  at a particular
position, the event positions (asterisks) are often 10\arcsec$-$20\arcsec\
from the event center.
Almost no events have been recorded near the active region loop systems due
to the confusion with loop brightenings.

\begin{figure}
   \includegraphics[width=8.5cm]{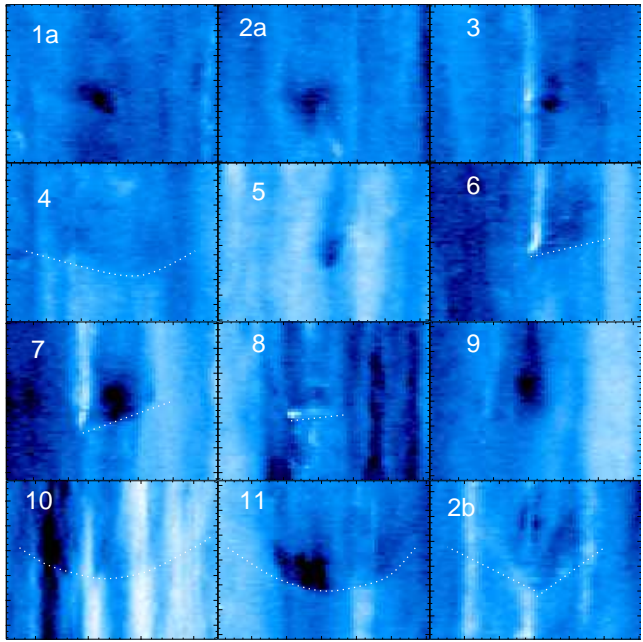}
   \caption{Close-up of events numbered in Fig.~\ref{space_time_fig}.
   The
   plane-of-sky velocities and lifetimes of
    the faint waves and dark features are given in Table~1.
White dotted lines trace the edges of the wave-like features. The frames are
100\arcsec\ by 125~min, except for event 10 which is 160\arcsec\ by 125~min.}
\label{mini_events_fig}
  \end{figure}

In Fig.~\ref{space_time_fig}, registered events have been circled.
Here a  variety of features have been
circled because different parts of events are crossing these particular
space-time slices. The larger ones are numbered and the measured lifetimes
and velocities in the sun-$x$ direction
of the dark and wave-like features are given in Table 1.
More detailed images of the individual events are shown in
Fig.~\ref{mini_events_fig}, where dotted lines are drawn along the edge of the
faint waves. These are sometimes curved because the waves
slow down. In these cases the velocity in the Table is the average velocity.
The lifetimes are minimum event lifetimes, since they
were measured from the displayed space-time images, and in a different direction
the lifetime may appear longer.
As can be seen in Fig.~\ref{space_time_fig}, for almost one third of the events it is not
possible to measure velocities. There are several darkenings that could be
events, but have not been included. This makes the statistics rather uncertain.
In the future we hope
to develop a more quantitative scheme.

The  1200\arcsec$\times1100$\arcsec\ area studied is mostly quiet Sun.
It covers about 1/8th of the solar surface. In this region, 176 events were recorded
in 24 hours.
This gives a rate of 1 event per min or 1400 per day over the whole Sun,
which is about a quarter of the rate of quiet Sun H$\alpha$ mini-filament eruptions found by
\citet{Wetal00}.
The typical lifetimes and velocities of the dark ejecta are very similar
to the H$\alpha$ mini-filaments so we believe these are the same phenomena.
The numbers in Table~1 indicate that the wave-like features have an average
velocity $\sim45$~\kms, and lifetime $\sim30$~min. Therefore  the distance travelled is
approximately 80~Mm.  Assuming the statistics of
Fig.~\ref{space_time_fig} are typical, one third of
events have detectable waves. This corresponds to 1/50th of the
lower corona being affected by these events at any one time.

\begin{table}
\caption{Velocity and lifetime of dark ejections and  faint waves in
Figs.~\ref{space_time_fig} and \ref{mini_events_fig}.}
\label{table1}
\centering
\begin{tabular}{c c c c} \\
\hline\hline
Event& Feature & V (km s$^{-1}$) & Lifetime (min) \\
\hline
1 & dark & 14 & 25 \\
2 & dark & 6 & 30 \\
  & wave & 26 & 40\\
3 & dark & 21 & 20\\
4 & wave & 41 & 19\\
5 & dark & 5 & 20\\
6 & dark & 19 & 45 \\
  & wave & 36 & 13 \\
7 & dark & 16 & 50 \\
& wave & 24 & 30\\
8 & wave & 100 & 6\\
9 & dark & 18 & 40 \\
10 & dark & 15 & 50 \\
   & wave& 50 & 30\\
11 & dark & 26 & 42 \\
    & wave & 36 & 60\\
    \hline
\end{tabular}
\end{table}

\section{Photospheric flows}
The activity of the corona is strongly coupled to the supergranular photospheric
flows \citep{Potts07}. It is well known that supergranular lanes and
the junctions of supergranular cells are associated with transition region
activity. Coupling the flows to the coronal emission was therefore a major
goal of this work.

After the basic flows have been computed using the Balltrack method with MDI
continuum images
\citep{Potts04}, we represent the photospheric transverse motion
 as arrows that
move with the flow.  New arrows are continually added at random positions
where there are
few arrows and removed from regions where they cluster together.
Arrows, acting like corks, collect at supergranular cell boundaries and
junctions, and disappear from cell centers.

\begin{figure*}
   \includegraphics[width=16cm]{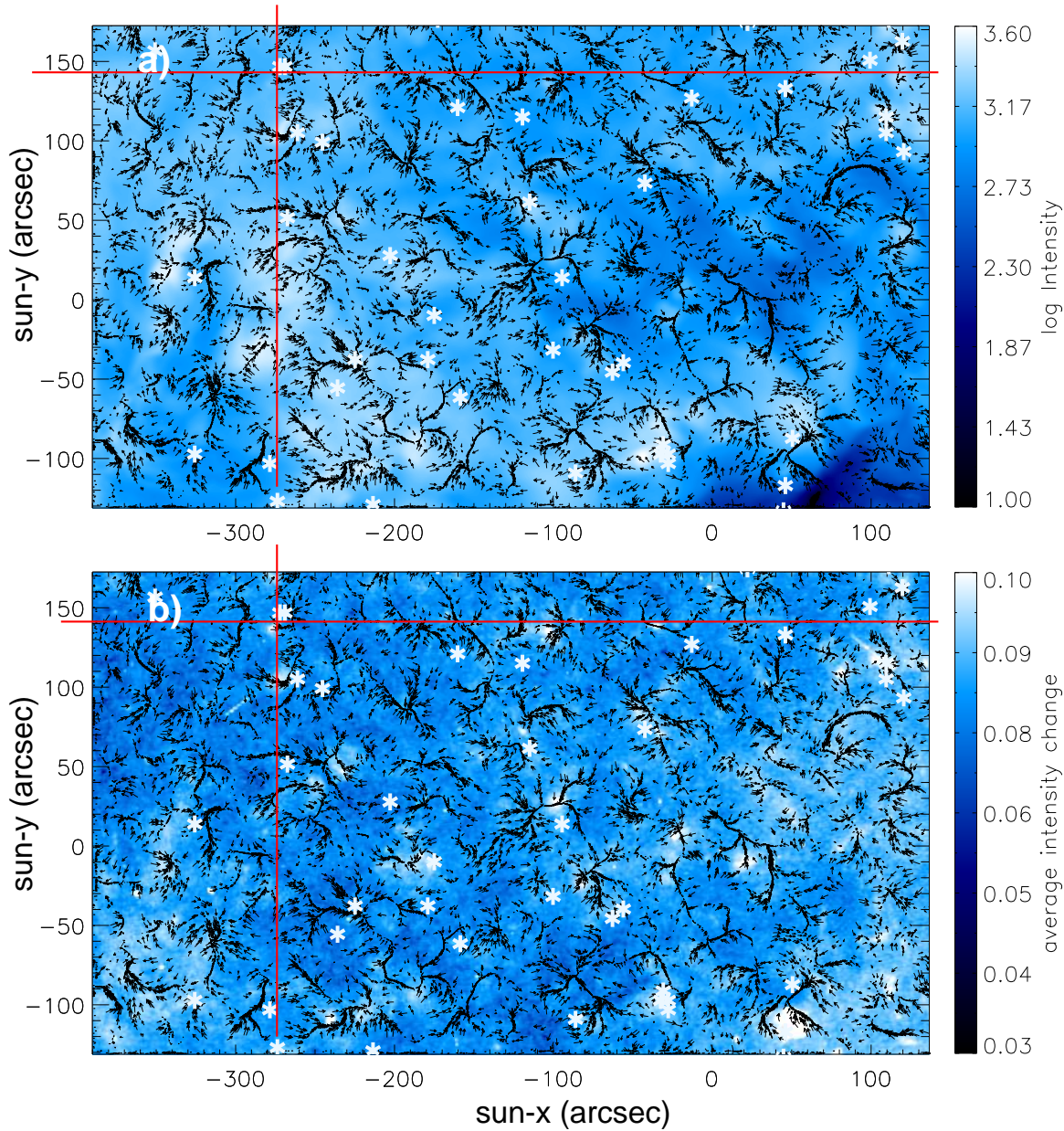}
   \caption{The relationship between photospheric flows, EUV intensity and
   brightenings, and the positions of mini-CME events.
   {\bf a)} The average 171~\AA\ intensity for the
   region studied on 11 June 2007.
{\bf b)} The average intensity change over 2.5~min (the time between images)
 normalized by the
average intensity.
The photospheric flows are represented as black arrows that accumulate along
the cell boundaries. The white asterisks indicate approximate positions
($\pm20$\arcsec) of mini-CME events.
The horizontal and vertical lines indicate the positions of the space-time images
in Fig.~\ref{space_time_fig}.
}
    \label{hrstevents_fig}
  \end{figure*}

For a single flow field, the arrows are integrated until  a converged solution
for that flow field is reached.
This representation of the average flow field over the 8 hour MDI period
is shown in Fig.~\ref{hrstevents_fig}, superimposed on the average 171~\AA\
intensity and the normalized average intensity change over 2.5~min.
The normalized average intensity change is the average of the running difference
images over the 8 hour period, divided by the average intensity.
When a time series of photospheric flows is available, then the arrow
positions respond to the changing flow field, so there is a small lag in the
arrow positions compared with the current flow field. For the discussion
here, where we are considering properties on the scale of supergranular cells,
 this lag is
insignificant. It ensures however that the arrows in the accompanying
animations follow smoothly from one frame to the next.

In Fig.~\ref{hrstevents_fig}, the positions of events are shown as white
asterisks.
 Events 1 and 2 in Fig.~\ref{space_time_fig} are the
two almost superimposed asterisks near (-270, 150).
The lower image shows that nearly all rapid intensity changes occur, as expected,
 along
supergranular boundaries, where the flows (represented by black arrows)
accumulate.
The biggest brightening occurred on the
 coronal hole boundary at (50,-120),
where there were two strong events.
Both were seen near the beginning of the 8 hour period so it was not possible
to see the photospheric flows in the build-up phase of these events. In the
next section we show the build-up and eruption  of two other events.

\subsection{Events}
As stated previously, there is a large variety of quiet Sun mini-CME events,
just like there is
a large variety of flares and CMEs. The characteristics are probably dependent
on structure of the event surroundings. In almost all cases both a
brightening and a dark plasmoid eruption were seen.
Here we show two events. One is a  small dark cloud eruption and the other
a flare-like event. Both were unique in their large-scale properties.

\begin{figure*}
\centering
   \includegraphics[width=15cm]{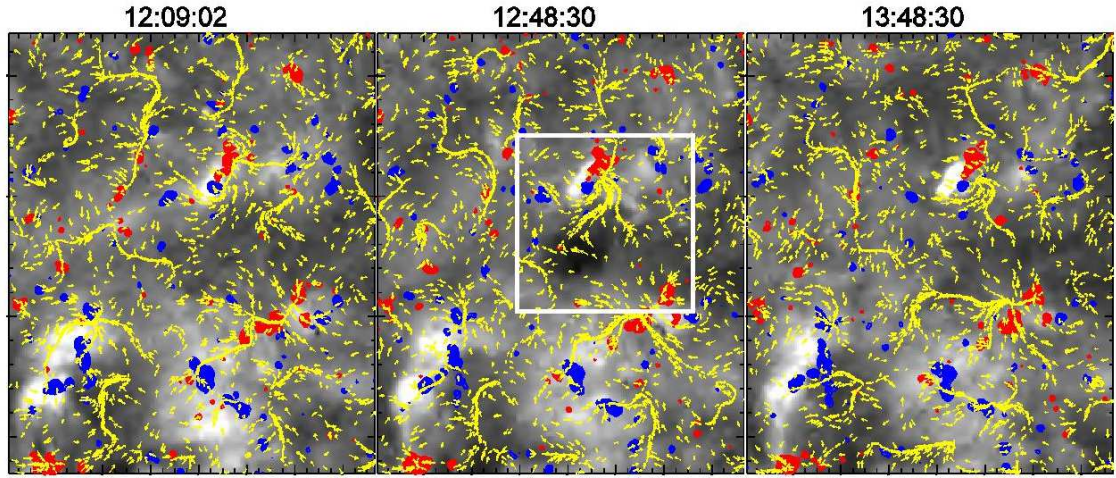}
   \caption{A cloud-like mini-CME. STEREO 171 images (greyscale) with
    regions of magnetic
   field greater (less) than 40~G (-40~G) coloured in blue (red), and the photospheric flows
   represented with yellow arrows.
   The FOV is 150\arcsec$\times180$\arcsec, centered at (-285,78). The white
   box outlines the event. A movie of the temporal evolution of the photospheric flows,
   magnetic field, and EUV intensity is shown in
 \href{http://www.mps.mpg.de/data/outgoing/innes/cmes/cloud_CME.gif}
 {\tt cloud\_CME.gif}}.
\label{ev34_fig}
  \end{figure*}

The first, shown in Fig.~\ref{ev34_fig} and in the animation
{\tt cloud\_CME.gif},
is the cloud-like eruption. Even in the quiet Sun there is a lot of activity
and events like these that appear as a darkening, are difficult to detect. We
have therefore put a white box around the event in the figure. The eruption
occurred from the outlined supergranular junction and extended to the region
of positive field on the right. The vortex in the photospheric flows
at the supergranulation junction is well established at the time of the
eruption.
After the dark cloud eruption, the region between the opposite polarity fields
brightened. It is interesting to note that the dark cloud seemed to stop at
the supergranular boundary to the south before fading into the background.

\begin{figure*}
\centering
   \includegraphics[width=16cm]{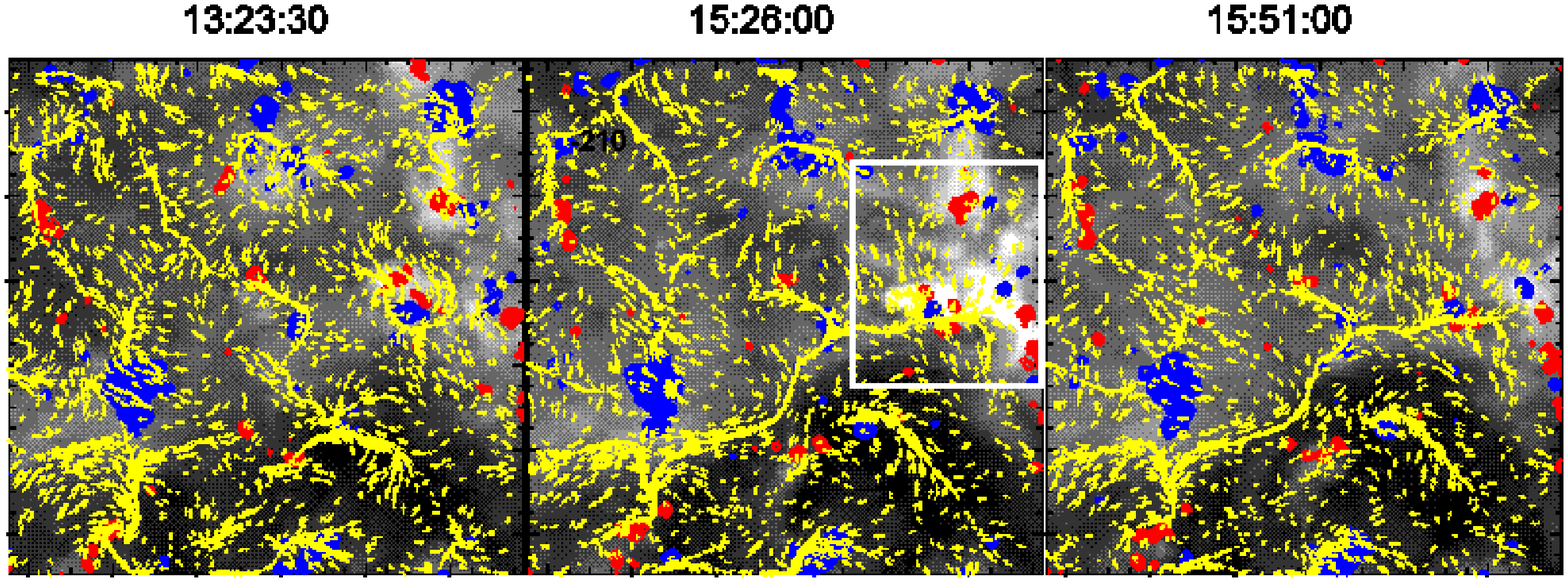}
  \caption{A flare-like mini-CME. The representation is the
   same as Fig.~\ref{ev34_fig} for
   FOV  125\arcsec$\times125$\arcsec,
    centered at (75,110). A movie of the photospheric flow, magnetic field, and
   EUV intensity is shown in
 \href{http://www.mps.mpg.de/data/outgoing/innes/cmes/flare_CME.gif}
 {\tt flare\_CME.gif},
 and of the EUV alone over a larger FOV in
 \href{http://www.mps.mpg.de/data/outgoing/innes/cmes/wave_CME.gif}
 {\tt wave\_CME.gif}.}
 \label{ev39_fig}
  \end{figure*}

The other event (Fig.~\ref{ev39_fig} and
in the animations {\tt flare\_CME.gif}
and {\tt wave\_CME.gif}),
 was a more spectacular quiet Sun mini-CME. It showed brightening,
filament eruption and wave-like dimming. The rotation seen in the photospheric
flows, at the
supergranular cell junction with opposite polarity fields,
started about three hours before the eruption.
After approximately 1.5
hours the flows changed quite significantly, producing a sharp lane flowing
eastward from the junction. It may have been this strong stretching of magnetic
footpoints that triggered the  eruption.
The event occurred on
the edge of the MDI field-of-view, so we were not able to trace the wave
propagation in relation to the supergranulation field structure. We show
the wave in the accompanying series of 171~\AA\ images, {\tt wave\_CME.gif},
moving very quickly to the top right corner. The
field-of-view in the animation is 150\arcsec$\times$150\arcsec.

\section{Discussion}
In this paper we have presented a preliminary overview of what we call
mini-CMEs in the quiet Sun. Most of the events found show both mini-filament
eruption and   microflare brightening, and in several events wave-like
features are seen propagating from the eruption site.
We have shown that in the two cases studied
the eruptions are driven by sheared
network core fields.
The natural question is whether these eruptions
 are responsible for coronal heating in the quiet Sun  as proposed
by \citet{MFPS99}. The event rate is 1400 per day over the whole Sun, but only
one third of these have observable wave-like fronts. If one
assumes a typical lifetime of 30~min and velocity of 45~\kms,
they are still too infrequent and small to affect a significant area of the
lower corona. Future observations from the Solar Dynamics Observatory (SDO)
should provide far superior images over a wider range of
temperatures as well as complementary  photospheric fields and flows, and
analysis of these images may provide further clues to how energy is transferred
to the corona.

\begin{acknowledgements}
We would like thank the referee for constructive comments. The data were
provided by the MDI/SOHO and EUVI/STEREO consortia.
STEREO is a project of NASA, SOHO a joint ESA/NASA project.
The EUVI data used here were produced by an international consortium of the
Naval Research Laboratory (USA), Lockheed Martin Solar and Astrophysics Lab
(USA), NASA Goddard Space Flight Center (USA),
Rutherford Appleton Laboratory (UK), University of Birmingham (UK),
Max-Planck-Institut for Solar System Research (Germany),
Centre Spatiale de Liège (Belgium), Institut d'Optique Théorique et Appliqueé (France),
and Institut d'Astrophysique Spatiale (France).
\end{acknowledgements}

\bibliographystyle{aa}


\end{document}